\documentclass[aps,prl,twocolumn,showpacs,lengthcheck,preprintnumbers]{revtex4}%
\usepackage{amsfonts}
\usepackage{amsmath}
\usepackage{amssymb}
\usepackage{graphicx}%
\begin{document}
\preprint{Phys. Rev. A \textbf{75}, 063610 (2007)}
\title{Ground-state properties of trapped Bose-Einstein condensates:\\Extension of the Thomas-Fermi approximation}
\author{A. Mu\~{n}oz Mateo}
\email{ammateo@ull.es}
\author{V. Delgado}
\email{vdelgado@ull.es}
\affiliation{Departamento de F\'{\i}sica Fundamental II, Universidad de La Laguna, La
Laguna, Tenerife, Spain}


\begin{abstract}
We derive general approximate formulas that provide with remarkable accuracy
the ground-state properties of any mean-field scalar Bose-Einstein condensate
with short-range repulsive interatomic interactions, confined in arbitrary
cylindrically symmetric harmonic traps. Our formulation is even applicable for
condensates containing a multiply quantized axisymmetric vortex. We have
checked the validity of our formulas by numerically solving the 3D
Gross-Pitaevskii equation.

\end{abstract}
\maketitle

%\author{A. Mu\~{n}oz Mateo and V. Delgado}

\section{I. INTRODUCTION}

Since the experimental realization of the first dilute Bose-Einstein
condensates (BECs) of trapped atomic gases \cite{BEC1,BEC2,BEC3} there has
been great interest in the study of the physical properties of ultracold
quantum gases \cite{RevStrin}. Under the usual experimental conditions, these
systems can be accurately described by the Gross-Pitaevskii equation (GPE)
\cite{GPE}, a mean-field equation of motion governing the behavior of the
condensate wave function $\psi(\mathbf{r})$. In the stationary case the GPE is
given by the nonlinear Schr\"{o}dinger equation
\begin{equation}
\left(  -\frac{\hbar^{2}}{2m}\nabla^{2}+V(\mathbf{r})+gN\left\vert
\psi\right\vert ^{2}\right)  \psi=\mu\psi,\label{eqGP}%
\end{equation}
where $\mu$ is the chemical potential, $N$ is the number of atoms,
$g=4\pi\hbar^{2}a/m$ is the interaction strength, determined by the
\textsl{s}-wave scattering length $a$, and $V(\mathbf{r})$ is the trap potential.

In this work we shall concentrate in the usual case of condensates with
repulsive interatomic interactions ($a>0$), confined in cylindrically
symmetric harmonic traps, $V(\mathbf{r})=\frac{1}{2}m(\omega_{\bot}^{2}%
r_{\bot}^{2}+\omega_{z}^{2}z^{2})$.

To obtain the condensate ground-state properties one has to solve the
nonlinear differential equation (\ref{eqGP}). No explicit analytical solutions
are known, so that, in general, this has to be done numerically. However,
solving numerically the 3D GPE is a nontrivial computational task, especially
for highly asymmetric trap geometries, where large basis or gridpoint sets can
be required to guarantee convergence. Fortunately, approximate analytical
solutions can be found in certain limiting cases. In the Thomas-Fermi (TF)
regime, which essentially occurs for condensates with a large number of atoms,
the kinetic energy can be neglected to a good approximation. In this case the
GPE reduces to a simple algebraic equation and one can obtain explicit
analytical expressions for the condensate ground-state properties
\cite{Baym1}. This approximation, however, cannot reproduce correctly the
decay of the wave function at the boundary of the atomic cloud, a region where
the kinetic energy has a decisive influence. Corrections to the TF
approximation have been proposed to account for the proper behavior of the
density cloud at the condensate surface \cite{Dalfovo1,Lundh1,Fet1}. In
the\ (ideal gas) perturbative regime, when the number of atoms in the
condensate is sufficiently small, explicit analytical expressions can also be
obtained by treating the interaction energy term in Eq. (\ref{eqGP}) as a weak perturbation.

Another approximation scheme that has proved to be useful\ in the
characterization of dilute BECs is that based on variational techniques
\cite{Baym1}. This approximation method has the additional interest that it
can be equally applied to the study of the condensate dynamics
\cite{Victor1,Reatto1}. By using appropriate variational trial wave functions,
the ground-state properties of trapped Bose-condensed gases have been studied
beyond the two analytically solvable regimes, for both isotropic \cite{Fet2}
and highly anisotropic condensates \cite{Das1}. For this purpose semiclassical
approximations ($\hbar\rightarrow0$) have also been considered
\cite{Vinas1,Huang1}. These approximate methods can be applied to the study of
vortex states as well \cite{Rok1,Sinha,Vinas2}.

Particular attention has been devoted to the physical properties of quasi-1D
\cite{Olsha1,Petrov1,Dunj1,Strin1,Reatto2} and quasi-2D \cite{Petrov2,Kett1}
condensates, systems so tightly confined in the radial or the axial dimension,
respectively, that the corresponding dynamics is restricted to zero-point oscillations.

In a previous work, by modifying the usual TF approximation conveniently, we
derived very accurate approximate analytical expressions for the ground-state
properties of trapped spherical, cigar-shaped, and disk-shaped condensates
with an arbitrary number of atoms in the mean-field regime \cite{Previo}. In
this work we extend our previous results and derive general approximate
formulas that provide with remarkable accuracy the ground-state properties of
any mean-field scalar Bose-Einstein condensate with short-range repulsive
interatomic interactions, confined in arbitrary cylindrically symmetric
harmonic traps, and even containing a multiply quantized axisymmetric vortex.

\section{II. MODEL}

We consider a BEC with an axisymmetric vortex line of topological charge $q$,
and confined in a harmonic trap characterized by oscillator lengths $a_{\bot
}=\sqrt{\hbar/m\omega_{\bot}}$ and $a_{z}=\sqrt{\hbar/m\omega_{z}}$. The trap
aspect ratio is given by $\lambda=\omega_{z}/\omega_{\bot}$. In this work we
shall distinguish between ground and vortex states only through the value of
the vortex charge. Accordingly, we shall refer to the lowest-energy state of
the condensate compatible with an axisymmetric vortex of charge $q$ as the
ground-state with a charge-$q$ vortex. This terminology will permit us to make
a unified treatment in which the actual ground state of the condensate is
simply a particular case corresponding to $q=0$.

For condensates with a sufficiently small number of atoms the mean-field
interaction energy can be treated as a weak perturbation. In this perturbative
regime the condensate wave function that minimizes the energy functional is
given, to the lowest order, by%
\begin{equation}
\psi_{q}(r_{\bot},z,\theta)=\exp(iq\theta)\varphi_{q}(r_{\bot})\phi
(z),\label{eq1}%
\end{equation}
with%
\begin{equation}
\varphi_{q}(r_{\bot})=(\pi a_{\bot}^{2}|q|!)^{-\frac{1}{2}}(r_{\bot}/a_{\bot
})^{|q|}\exp(-r_{\bot}^{2}/2a_{\bot}^{2}),\label{eq2a}%
\end{equation}%
\begin{equation}
\phi(z)=(\pi a_{z}^{2})^{-1/4}\exp(-z^{2}/2a_{z}^{2}).\label{eq2b}%
\end{equation}
Using this result one obtains the following analytical expression for the
chemical potential:%
\begin{equation}
\mu=\frac{1}{2}\hbar\omega_{z}+(|q|+1)\hbar\omega_{\bot}+g\bar{n},\label{eq3}%
\end{equation}
where $\bar{n}=c_{q}N/(2\pi)^{3/2}a_{\bot}^{2}a_{z}$ is the mean atom density
and the coefficient $c_{q}$ is a function of the vortex charge that takes
values in the interval $[1,0)$ and accounts for the dilution effect that the
centrifugal force associated with the vortex has on the condensate mean
density%
\begin{equation}
c_{q}=\frac{(2|q|)!}{2^{2|q|}(|q|!)^{2}}.\label{eq4}%
\end{equation}
For $q=0$ one has $c_{q}=1$, and the usual results for the ground-state
properties of a BEC in the perturbative regime are recovered. For a unit
charge vortex the dilution coefficient\ takes the value $1/2$, and, in
general, for $|q|\gg1$, it decreases slowly as $1/\sqrt{\pi|q|}$.

In the Thomas-Fermi regime, for condensates in the ground state $(q=0)$ and
with a sufficiently large number of atoms, one can neglect the kinetic energy
in comparison with the interaction energy, and the stationary GPE leads to%
\begin{equation}
\mu=\frac{1}{2}m\omega_{z}^{2}z^{2}+\frac{1}{2}m\omega_{\bot}^{2}r_{\bot}%
^{2}+gN\left\vert \psi(r_{\bot},z)\right\vert ^{2}. \label{eq5}%
\end{equation}

In the presence of a vortex of charge $q\neq0$ the above equation is no longer
a good approximation. In this case neglecting the kinetic energy amounts to
neglecting the vortex itself. However, for a large $N$, the condensate density
cloud outside the vortex core is still given, to a very good approximation, by
the Thomas-Fermi profile. We thus shall assume the above TF expression to be
valid up to a lower cutoff radius%
\begin{equation}
r_{\bot}^{0}=\sqrt{2(|q|+1)}a_{\bot},\label{eq5b}%
\end{equation}
determined from the condition that, in the presence of a vortex, the
contribution from the radial harmonic oscillator energy cannot be smaller than
$(|q|+1)\hbar\omega_{\bot}$, i.e.,%
\begin{equation}
\frac{1}{2}m\omega_{\bot}^{2}(r_{\bot}^{0})^{2}=(|q|+1)\hbar\omega_{\bot
}.\label{eq6}%
\end{equation}
Likewise, the contribution from the axial harmonic oscillator energy
should\ not be smaller than the corresponding zero-point energy. To account
for this fact we introduce a second cutoff $z_{0}$, defined through the
condition%
\begin{equation}
\frac{1}{2}m\omega_{z}^{2}z_{0}^{2}=\frac{1}{2}\hbar\omega_{z},\label{eq7}%
\end{equation}
which yields the axial cutoff%
\begin{equation}
z_{0}=a_{z}.\label{eq7b}%
\end{equation}
This leads us to consider the TF expression (\ref{eq5}) applicable in a volume
$V_{3}$ that corresponds to the usual TF ellipsoidal density cloud truncated
at both $r_{\bot}=r_{\bot}^{0}$ and $\,|z|=z_{0}$,%
\begin{multline}
V_{3}\equiv\left\{  (r_{\bot},z)\colon\,r_{\bot}^{2}/R_{\mathrm{TF}}^{2}%
+z^{2}/Z_{\mathrm{TF}}^{2}\leq1,\right.  \\
\left.  r_{\bot}>r_{\bot}^{0}\,\;\wedge\;|z|>z_{0}\right\}  .\label{eq8}%
\end{multline}

In the above equation $R_{\mathrm{TF}}\equiv\sqrt{2\mu/\hbar\omega_{\bot}%
}\,a_{\bot}$ and $Z_{\mathrm{TF}}\equiv\sqrt{2\mu/\hbar\omega_{z}}\,a_{z}$ are
the usual TF values for the condensate radius and axial half-length,
respectively. In our approach, however, the condensate radius and axial
half-length do not coincide with the above TF expressions. Because of the
cutoffs we have introduced they are given instead by%
\begin{equation}
R=\sqrt{2\mu/\hbar\omega_{\bot}-\lambda}\;a_{\bot}, \label{eq9}%
\end{equation}%
\begin{equation}
Z=\sqrt{2\mu/\hbar\omega_{z}-2(|q|+1)/\lambda}\;a_{z}. \label{eq10}%
\end{equation}
However, when $\mu\gg(|q|+1)\hbar\omega_{\bot}$, then $Z$ becomes
indistinguishable from $Z_{\mathrm{TF}}$. This is true, in particular,
whenever $\lambda\gg2(|q|+1)$. Likewise, if $\mu\gg\frac{1}{2}\hbar\omega_{z}%
$, then $R\rightarrow R_{\mathrm{TF}}$.

More generally, for condensates with $\mu\gg(|q|+1)\hbar\omega_{\bot}$, which
occurs whenever $\lambda\gg2(|q|+1)$, the relative contribution to the
condensate properties coming from the region $\,r_{\bot}\leq r_{\bot}^{0}$
becomes negligible. The same occurs with the relative contribution from the
region corresponding to $|z|\leq z_{0}$ when the chemical potential is much
larger than the axial zero-point energy, which occurs whenever $\lambda\ll1$.
It is then clear that when the number of atoms is sufficiently large that
$\mu\gg(|q|+1)\hbar\omega_{\bot}+\frac{1}{2}\hbar\omega_{z}$ the dominant
contribution comes from the $V_{3}$ volume introduced above.

Only for condensates with $\mu\simeq(|q|+1)\hbar\omega_{\bot}$ is the
contribution from the region $r_{\bot}\leq r_{\bot}^{0}$\ the most significant
one. These are condensates with the transverse dynamics frozen in the lowest
energy state compatible with a charge-$q$ axisymmetric vortex. In this case
the condensate wave function can be factorized as $\psi_{q}(r_{\bot}%
,z,\theta)=\exp(iq\theta)\varphi_{q}(r_{\bot})\phi(z)$, with $\varphi
_{q}(r_{\bot})$ given by Eq. (\ref{eq2a}). After substituting this wave
function in the stationary GPE, multiplying by $\psi_{q}^{\ast}$, and
integrating over the radial dynamics, one obtains%
\begin{equation}
\frac{1}{2}m\omega_{z}^{2}z^{2}+(|q|+1)\hbar\omega_{\bot}+g_{\mathrm{1D}%
}N|\phi(z)|^{2}=\mu,\label{eq11}%
\end{equation}
where $g_{\mathrm{1D}}=c_{q}g/2\pi a_{\bot}^{2}$, and we have neglected the
axial kinetic energy ($\sim\frac{1}{2}\hbar\omega_{z}$) against $(|q|+1)\hbar
\omega_{\bot}$. It is convenient to rewrite $g_{\mathrm{1D}}$ as $g\kappa
_{1}^{-1}\bar{n}_{2}$, where $\bar{n}_{2}=1/\pi(r_{\bot}^{0})^{2}=1/[2\pi
a_{\bot}^{2}(|q|+1)]$\ is a uniform mean density per unit area normalized to
unity in the region $r_{\bot}\leq r_{\bot}^{0}$, and%
\begin{equation}
\kappa_{1}^{-1}\equiv(|q|+1)c_{q}\label{eq12}%
\end{equation}
is the appropriate renormalization factor. The important point is that the
effect of the vortex can be incorporated in an exact manner into a localized
uniform mean density $\bar{n}_{2}$ with a renormalized interaction strength
$g\kappa_{1}^{-1}$. We shall assume that, to a good approximation, this still
remains true for condensates with an arbitrary chemical potential.

On the other hand, the contribution from the region $|z|\leq z_{0}$ is the
most significant one only for condensates with $\mu\simeq\frac{1}{2}%
\hbar\omega_{z}$. In this case the axial dynamics is restricted to zero-point
oscillations, and the wave function of such quasi-2D condensates can be
written as $\psi_{q}(r_{\bot},z,\theta)=\exp(iq\theta)\varphi(r_{\bot}%
)\phi(z)$, where now $\phi(z)$ is given by Eq. (\ref{eq2b}). Substituting into
the stationary GPE and integrating out the axial dynamics one obtains%
\begin{equation}
\frac{1}{2}\hbar\omega_{z}+\frac{1}{2}m\omega_{\bot}^{2}r_{\bot}%
^{2}+g_{\mathrm{2D}}N|\varphi(r_{\bot})|^{2}=\mu,\label{eq13}%
\end{equation}
where $g_{\mathrm{2D}}=g/\sqrt{2\pi}\,a_{z}$, and now we have neglected the
radial kinetic energy against the axial zero-point energy, which is a good
approximation as long as $\mu\simeq\frac{1}{2}\hbar\omega_{z}$. As before, it
is convenient to rewrite $g_{\mathrm{2D}}$ in terms of a uniform mean density
per unit length normalized to unity in the volume $|z|\leq z_{0}$. To this end
we introduce a renormalization factor $\kappa_{2}^{-1}\equiv\sqrt{2/\pi}$ and
rewrite $g_{\mathrm{2D}}$ as $g\kappa_{2}^{-1}\bar{n}_{1}$ with $\bar{n}%
_{1}=1/2a_{z}$. This indicates that the contribution from the axial zero-point
oscillations, which is the dominant contribution in these quasi-2D
condensates, can be properly accounted for by simply introducing a localized
uniform mean density per unit length with a renormalized interaction strength.
Again, we shall assume this to be also valid for condensates with an arbitrary
chemical potential. One finds, however, that somewhat more accurate results
are obtained when one lets the renormalization factor $\kappa_{2}^{-1}$ to
approach unity in the TF regime \cite{Previo}. Since the final results are
little sensitive to the specific functional form of $\kappa_{2}^{-1}$, we
propose one of the simplest possibilities \cite{Previo}%

\begin{align}
\kappa_{2}^{-1}(\chi_{2})  &  \equiv\sqrt{2/\pi}+\Theta(\chi_{2}%
-0.1)\nonumber\\
\times &  \left(  1-\sqrt{2/\pi}\right)  \left(  1-\frac{R_{\mathrm{TF}}%
(\chi_{2}=0.1)}{R_{\mathrm{TF}}(\chi_{2})}\right)  , \label{eq14}%
\end{align}
where $\Theta(x)$ is the step function, $\chi_{2}\equiv Na/\lambda^{2}a_{z}$,
and $R_{\mathrm{TF}}(\chi_{2})=(15\chi_{2})^{1/5}a_{\bot}$ is the TF radius.

Motivated by the above ideas and the fact that there exists a direct relation
between the number of atoms and the size of a trapped BEC, we propose the
following ansatz for the ground-state properties of any mean-field scalar
Bose-Einstein condensate with an axisymmetric vortex of charge $q$, confined
in an arbitrary axisymmetric harmonic trap:%
\begin{align*}
\frac{1}{2}m\omega_{z}^{2}z^{2}+\frac{1}{2}m\omega_{\bot}^{2}r_{\bot}%
^{2}+gN\left\vert \psi(r_{\bot},z)\right\vert ^{2} &  =\mu,\hspace
{0.5cm}\mathbf{r}\in V_{3}\\
\frac{1}{2}\hbar\omega_{z}+\frac{1}{2}m\omega_{\bot}^{2}r_{\bot}^{2}%
+g\kappa_{2}^{-1}N\bar{n}_{1}|\varphi(r_{\bot})|^{2} &  =\mu,\hspace
{0.5cm}\mathbf{r}\in V_{2}\\
\frac{1}{2}m\omega_{z}^{2}z^{2}+\alpha_{q}\hbar\omega_{\bot}+g\kappa_{1}%
^{-1}N\bar{n}_{2}|\phi(z)|^{2} &  =\mu,\hspace{0.5cm}\mathbf{r}\in V_{1}\\
\frac{1}{2}\hbar\omega_{z}+\alpha_{q}\hbar\omega_{\bot}+g\kappa_{0}^{-1}%
\bar{n}_{0} &  =\mu,\hspace{0.5cm}\mathbf{r}\in V_{0}%
\end{align*}
with $\psi=0$ elsewhere. In the above equations $\alpha_{q}\equiv(|q|+1)$ and
$\kappa_{0}\equiv\kappa_{1}\kappa_{2}$, with $\kappa_{1}$ and $\kappa_{2}$
defined by Eqs. (\ref{eq12}) and (\ref{eq14}), respectively. As already seen,
$\bar{n}_{1}=1/2a_{z}$ and $\bar{n}_{2}=1/2\pi a_{\bot}^{2}\alpha_{q}$,
whereas $\bar{n}_{0}$ is an effective mean density (per unit volume) localized
in $V_{0}$ and defined in terms of $\mu$ through the above expressions. Note
that $\kappa_{0}^{-1}$ is exactly the renormalization constant required to
make the latter of the above equations compatible with the other ones as well
as with the perturbative result (\ref{eq3}). The outer volume $V_{3}$ is
defined by Eq. (\ref{eq8}) while the remaining inner volumes are defined by%
\[
V_{2}\equiv\left\{  (r_{\bot},z)\colon\,r_{\bot}^{0}<\,r_{\bot}\leq
R\,\;\wedge\;|z|\leq z_{0}\right\}  ,
\]%
\[
V_{1}\equiv\left\{  (r_{\bot},z)\colon\,r_{\bot}\leq r_{\bot}^{0}%
\,\;\wedge\;z_{0}<|z|\leq Z\right\}  ,
\]%
\[
V_{0}\equiv\left\{  (r_{\bot},z)\colon\,r_{\bot}\leq r_{\bot}^{0}%
\,\;\wedge\;|z|\leq z_{0}\right\}  .
\]

The ansatz we have just introduced represents a direct generalization of our
previous proposal in Ref. \cite{Previo} and extends the applicability of the
approach to mean-field condensates confined in axisymmetric harmonic traps
with an arbitrary geometry, and containing an axisymmetric vortex of charge
$q$. As we shall see, in the appropriate limits the results obtained in the
present work reduce to those previously obtained in Ref. \cite{Previo}.

From Eq. (\ref{eq10}) the chemical potential can be written in terms of the
dimensionless axial half-length $\overline{Z}\equiv Z/a_{z}$ as%
\begin{equation}
\frac{\mu}{\hbar\omega_{z}}=\frac{1}{2}\overline{Z}^{2}+\frac{1}{\lambda
}(|q|+1).\label{eq15}%
\end{equation}
As usual, the condition that the condensate contains $N$ particles determines
the precise value of $\mu$. After a straightforward calculation one obtains%
\begin{align}
\frac{\chi_{0}}{\lambda^{5/3}} &  =\frac{1}{15}\overline{Z}^{5}+\frac{\xi_{1}%
}{8}\overline{Z}^{4}+\frac{\beta_{q}}{3\lambda}\overline{Z}^{3}\nonumber\\
&  +\frac{1}{2}\left(  \frac{\beta_{q}\xi_{1}}{\lambda}-\frac{\xi_{3}}%
{2}\right)  \overline{Z}^{2}-\frac{1}{2}\left(  \frac{\beta_{q}\xi_{3}%
}{\lambda}-\frac{\xi_{5}}{4}\right)  ,\label{eq16}%
\end{align}
where $\beta_{q}\equiv\kappa_{1}(|q|+1)=1/c_{q}$, $\xi_{n}\equiv(\kappa
_{2}-1/n)$ with $n=1,3,5,\ldots$, and we have defined the dimensionless
interaction parameter $\chi_{0}$ as%
\begin{equation}
\chi_{0}\equiv Na/a_{0},\label{eq17}%
\end{equation}
with $a_{0}\equiv(a_{\bot}^{2}a_{z})^{1/3}$ being the mean oscillator length.
From Eqs. (\ref{eq9}) and (\ref{eq10}) one also finds the following expression
for the condensate radius $\overline{R}\equiv R/a_{\bot}$:%
\begin{equation}
\overline{R}^{2}=\lambda(\overline{Z}^{2}-1)+2(|q|+1).\label{eq18}%
\end{equation}

The mean-field interaction energy per particle $\epsilon_{\mathrm{int}}\equiv
E_{\mathrm{int}}/N$ is defined by $\epsilon_{\mathrm{int}}=(1/2)\int
g_{r}N\left\vert \psi(\mathbf{r})\right\vert ^{4}d^{3}\mathbf{r}$, where
$N\left\vert \psi\right\vert ^{2}$ represents the local density in each region
and $g_{r}$ denotes the corresponding renormalized interaction strength. After
some calculation one obtains%
\begin{align}
\frac{\epsilon_{\mathrm{int}}}{\hbar\omega_{z}} &  =\frac{\lambda^{5/3}}%
{8\chi_{0}}\left[  \frac{8}{105}\overline{Z}^{7}+\frac{\xi_{1}}{6}\overline
{Z}^{6}+\frac{8\beta_{q}}{15\lambda}\overline{Z}^{5}\right.  \nonumber\\
&  +\left(  \frac{\beta_{q}\xi_{1}}{\lambda}-\frac{\xi_{3}}{2}\right)
\overline{Z}^{4}-2\left(  \frac{\beta_{q}\xi_{3}}{\lambda}-\frac{\xi_{5}}%
{4}\right)  \overline{Z}^{2}\nonumber\\
&  \left.  +\left(  \frac{\beta_{q}\xi_{5}}{\lambda}-\frac{\xi_{7}}{6}\right)
\right]  .\label{eq19}%
\end{align}

Finally, the kinetic and potential energies can be readily obtained in terms
of the previous results by using the \emph{exact} relations \cite{RevStrin}
\begin{subequations}
\label{eq20}%
\begin{align}
\epsilon_{\mathrm{kin}}  &  \equiv E_{\mathrm{kin}}/N=\mu
/2-(7/4)E_{\mathrm{int}}/N,\label{eq20a}\\
\epsilon_{\mathrm{pot}}  &  \equiv E_{\mathrm{pot}}/N=\mu
/2-(1/4)E_{\mathrm{int}}/N. \label{eq20b}%
\end{align}

Equations (\ref{eq15})--(\ref{eq20}) provide the ground-state properties we
are looking for. All that is needed is to solve the quintic polynomial
equation (\ref{eq16}). This is a general equation that provides the axial
half-length $\overline{Z}$ of any mean-field scalar condensate as a function
of only three parameters: the interaction parameter $\chi_{0}$, the trap
aspect ratio $\lambda$, and the vortex charge $q$. In certain particular cases
it is possible to find useful approximate analytical solutions. However, in
general, Eq. (\ref{eq16}) has to be solved numerically. It is important to
note that this is a trivial computational task that can be done immediately
with the built-in capabilities of symbolic computational software packages
such as \textsc{mathematica} or \textsc{matlab}. In fact, to obtain the roots
of a polynomial one simply has to type in a single instruction and the answer
is instantaneous.

\section{III. LIMITING CASES}

The above formulas simplify considerably in two limiting cases that,
essentially, correspond to condensates confined in disk-shaped traps
satisfying $\lambda\gg2(|q|+1)$ and cigar-shaped traps satisfying $\lambda
\ll1$. We have already found these two limiting cases before. As mentioned
above, in the first case the relative contribution to the condensate
properties coming from the inner cylinder $\,r_{\bot}\leq r_{\bot}^{0}$
becomes negligible, while, in the second case, it is the relative contribution
from the inner disk $|z|\leq z_{0}$ that becomes negligible. Under these
circumstances we shall be able to find approximate analytical solutions of the
polynomial equation (\ref{eq16}).

\subsection{Disk-shaped traps}

Taking into account that $\beta_{q}/\lambda\leq(|q|+1)/\lambda$, it follows
that in the limit $\lambda\gg2(|q|+1)$ Eq. (\ref{eq16}) reduces to%
\end{subequations}
\begin{equation}
\chi_{2}=\frac{1}{15}\overline{Z}^{5}+\frac{\xi_{1}}{8}\overline{Z}^{4}%
-\frac{\xi_{3}}{4}\overline{Z}^{2}+\frac{\xi_{5}}{8},\label{eq21}%
\end{equation}
where $\chi_{2}\equiv Na/\lambda^{2}a_{z}$ is now the only relevant physical
parameter. Using Eq. (\ref{eq18}) one can easily see that for $q=0$ the above
equation coincides exactly with that obtained previously in Ref.
\cite{Previo}. This is true in general (i.e., for any $q$) whenever
$\lambda\gg2(|q|+1)$. Under these circumstances the contribution of the vortex
can be neglected to a good approximation and we can use the analytical
solution found in Ref. \cite{Previo}%
\begin{equation}
\overline{Z}^{2}=1+\left[  \left(  1/15\chi_{2}\right)  ^{8/5}+\left(
\kappa_{2}/8\chi_{2}\right)  ^{2}\right]  ^{-1/4}.\label{eq22}%
\end{equation}
From this result one immediately obtains the chemical potential using Eq.
(\ref{eq15})%
\begin{equation}
\frac{\mu}{\hbar\omega_{z}}=\frac{1}{2}\overline{Z}^{2}.\label{eq22b}%
\end{equation}

On the other hand, in the limit we are considering, the interaction energy
(\ref{eq19})\ becomes%
\begin{equation}
\frac{\epsilon_{\mathrm{int}}}{\hbar\omega_{z}}=\frac{1}{8\chi_{2}}\left(
\frac{8}{105}\overline{Z}^{7}+\frac{\xi_{1}}{6}\overline{Z}^{6}-\frac{\xi_{3}%
}{2}\overline{Z}^{4}+\frac{\xi_{5}}{2}\overline{Z}^{2}-\frac{\xi_{7}}%
{6}\right)  .\label{eq23}%
\end{equation}
As before, it is not hard to see that this equation is the same as that
obtained previously\ in Ref. \cite{Previo}.

Another relevant physical quantity in the characterization of disk-shaped
condensates is the condensate density per unit area, defined as $n_{2}%
(r_{\bot})\equiv N\int dz\left\vert \psi(r_{\bot},z)\right\vert ^{2}$. Outside
the vortex core ($r_{\bot}>r_{\bot}^{0}$), which we are neglecting in this
limit, a straightforward calculation leads to%
\begin{equation}
n_{2}(r_{\bot})=\frac{\xi_{1}\left[  2\overline{\mu}_{z}(r_{\bot})-1\right]
}{4\pi aa_{z}}+\frac{\left[  2\overline{\mu}_{z}(r_{\bot})\right]  ^{3/2}%
-1}{6\pi aa_{z}},\label{eq23b}%
\end{equation}
where $n_{2}(r_{\bot}>R)=0$ and $\overline{\mu}_{z}(r_{\bot})\equiv\mu
_{z}(r_{\bot})/\hbar\omega_{z}$ is given by%
\begin{equation}
\overline{\mu}_{z}(r_{\bot})\equiv\frac{1}{2}+\frac{1}{2}\left(  \overline
{R}/\sqrt{\lambda}\right)  ^{2}\left(  1-\frac{r_{\bot}^{2}}{R^{2}}\right)
,\label{eq23c}%
\end{equation}
with $\overline{R}^{2}\simeq\lambda(\overline{Z}^{2}-1)$ [Eq. (\ref{eq18})].
Again, this result coincides with that derived in Ref. \cite{Previo}. In fact,
in general, in the limit $\lambda\gg2(|q|+1)$ the formalism proposed in the
previous section becomes independent of both $\lambda$ and $q$ and reduces to
that developed in Ref. \cite{Previo}.

It can be easily verify that $\mu_{z}(r_{\bot})$ is nothing but the local
chemical potential, defined as $\mu_{z}(r_{\bot})=\mu-\frac{1}{2}m\omega
_{\bot}^{2}r_{\bot}^{2}$ \cite{Huang1}. Equation (\ref{eq23b}) is a cubic
equation in $\overline{\mu}_{z}^{1/2}$ which has only one real solution.
Solving this equation, after some algebra one finds the following expression
for the local chemical potential as a function of the condensate density per
unit area \cite{Abramo}:%
\begin{equation}
\overline{\mu}_{z}(r_{\bot})\equiv{}\!\frac{1}{8}\!\left[  \!\left(
\!\eta+\!\sqrt{\eta^{2}\!-\!\xi_{1}^{6}}\right)  ^{\frac{1}{3}}\!+\!\left(
\!\eta\!-\!\sqrt{\eta^{2}\!-\!\xi_{1}^{6}}\right)  ^{\frac{1}{3}}%
\!\!-\!\xi_{1}\right]  ^{2},\label{eq23d}%
\end{equation}
where $\eta=4+6\xi_{1}-\xi_{1}^{3}+24\pi aa_{z}n_{2}(r_{\bot})$. In the TF
regime ($\chi_{2}\gg1\rightarrow aa_{z}n_{2}\gg1$), the above equation reduces
to%
\begin{equation}
\overline{\mu}_{z}(r_{\bot})=\left[  (3\pi/\sqrt{2})aa_{z}n_{2}(r_{\bot
})\right]  ^{2/3},\label{eq23e}%
\end{equation}
which coincides with the expression that can be obtained directly from the 3D
Gross-Pitaevskii equation in this regime. In the quasi-2D perturbative limit
($\chi_{2}\ll1\rightarrow aa_{z}n_{2}\ll1$), Eq. (\ref{eq23d}) reduces to%
\begin{equation}
\overline{\mu}_{z}(r_{\bot})=1/2+2\sqrt{2\pi}aa_{z}n_{2}(r_{\bot
}).\label{eq23f}%
\end{equation}
This is again the correct result, as follows from the perturbative solution of
the GPE in this limit.

Equations (\ref{eq23b})--(\ref{eq23d}) permit us to derive a general formula
for the (local) radial (first) sound velocity $c_{\mathrm{2D}}$ of a
disk-shaped condensate, which is defined by%
\begin{equation}
c_{\mathrm{2D}}^{2}=\frac{n_{2}}{m}\frac{\partial\mu_{z}}{\partial n_{2}%
}.\label{eq23g}%
\end{equation}
From Eq. (\ref{eq23b}) one immediately obtains%
\begin{equation}
\frac{mc_{\mathrm{2D}}^{2}}{\hbar\omega_{z}}=\frac{\frac{3}{2}\xi_{1}\left[
2\overline{\mu}_{z}(r_{\bot})-1\right]  +\left[  2\overline{\mu}_{z}(r_{\bot
})\right]  ^{3/2}-1}{3\xi_{1}+3\left[  2\overline{\mu}_{z}(r_{\bot})\right]
^{1/2}}.\label{eq23h}%
\end{equation}
This equation in the TF regime ($\chi_{2}\gg1$) reduces to%
\begin{equation}
\frac{mc_{\mathrm{2D}}^{2}}{\hbar\omega_{z}}=\frac{2}{3}\overline{\mu}%
_{z}(r_{\bot})=\left(  \frac{2\pi}{\sqrt{3}}aa_{z}n_{2}(r_{\bot})\right)
^{2/3},\label{eq23i}%
\end{equation}
while in the quasi-2D perturbative regime ($\chi_{2}\ll1$), it reduces to%
\begin{equation}
\frac{mc_{\mathrm{2D}}^{2}}{\hbar\omega_{z}}=\overline{\mu}_{z}(r_{\bot
})-\frac{1}{2}=2\sqrt{2\pi}aa_{z}n_{2}(r_{\bot}).\label{eq23j}%
\end{equation}
These are the correct limits as follows from the substitution of Eqs.
(\ref{eq23e}) and (\ref{eq23f}) into Eq. (\ref{eq23g}). In fact, it is not
hard to see that all the analytical formulas derived in this section reduce to
the correct expressions in both the TF and the perturbative regimes
\cite{Previo}.

For the particular case of a homogeneous disk-shaped condensate (no radial
confinement and $n_{2}$ constant) the authors of Ref. \cite{Reatto2} obtained
an expression for the radial sound velocity that interpolates between the
above two regimes. Such expression leads to the correct perturbative result
and reproduces to a good approximation the TF result. Even though this
expression is analytic, it is too complicated to be written in a useful
compact way and the authors of Ref. \cite{Reatto2} do not provide an explicit
analytical formula in their work.

\subsection{Cigar-shaped traps}

In the $\lambda\ll1$ limit Eq. (\ref{eq16}) reduces to%
\begin{equation}
\chi_{1}=\frac{1}{15}(\sqrt{\lambda}\,\overline{Z})^{5}+\frac{1}{3}\beta
_{q}(\sqrt{\lambda}\,\overline{Z})^{3}, \label{eq24}%
\end{equation}
with $\chi_{1}\equiv\lambda Na/a_{\bot}$. In this limit the formalism becomes
independent of $\lambda$ and the vortex contribution enters in a rather simple
way through the parameter%
\begin{equation}
\beta_{q}=\frac{2^{2|q|}(|q|!)^{2}}{(2|q|)!}. \label{eq24bis}%
\end{equation}
This fact will permit us to find an approximate analytical solution.

Given a polynomial equation $P(x)=\chi$, we define the residual error
associated with the approximate solution $x_{\varepsilon}$ as
$(P(x_{\varepsilon})-\chi)/\chi$ \cite{Previo}. We have explicitly verified
that the expression%
\begin{equation}
\sqrt{\lambda}\,\overline{Z}=\left[  \frac{1}{\left(  15\chi_{1}\right)
^{\frac{4}{5}}+\frac{1}{3}}+\frac{1}{57\chi_{1}+345}+\frac{1}{(3\chi_{1}%
/\beta_{q})^{\frac{4}{3}}}\right]  ^{-\frac{1}{4}}\label{eq24b}%
\end{equation}
satisfies Eq. (\ref{eq24}) with a residual error smaller than $3.2\%$ for any
$\chi_{1}\in\lbrack0,\infty)$ and $0\leq|q|\leq10$. In fact, as seen in Ref.
\cite{Previo}, in the absence of vortices $(q=0)$ the above solution is
somewhat more accurate and the error is less than $0.75\%$.

From Eq. (\ref{eq15}) the chemical potential is given now by%
\begin{equation}
\frac{\mu}{\hbar\omega_{\bot}}=(|q|+1)+\frac{1}{2}(\sqrt{\lambda}%
\,\overline{Z})^{2}.\label{eq24c}%
\end{equation}
In the $\lambda\ll1$ limit the interaction energy\ becomes%
\begin{equation}
\frac{\epsilon_{\mathrm{int}}}{\hbar\omega_{\bot}}=\frac{1}{15\chi_{1}}\left[
\frac{1}{7}(\sqrt{\lambda}\,\overline{Z})^{7}+\beta_{q}(\sqrt{\lambda
}\,\overline{Z})^{5}\right]  .\label{eq25}%
\end{equation}
On the other hand, the condensate density per unit length, $n_{1}(z)\equiv
N\int2\pi r_{\bot}dr_{\bot}\left\vert \psi(r_{\bot},z)\right\vert ^{2}$, is%
\begin{equation}
n_{1}(z)=\beta_{q}\frac{(\sqrt{\lambda}\;\overline{Z})^{2}}{4a}\left(
1-\frac{z^{2}}{Z^{2}}\right)  +\frac{(\sqrt{\lambda}\;\overline{Z})^{4}}%
{16a}\left(  1-\frac{z^{2}}{Z^{2}}\right)  ^{2}\label{eq26}%
\end{equation}
with $n_{1}(z)=0$ for $|z|>Z$.

The local chemical potential, defined as $\mu_{\bot}(z)\equiv\mu-\frac{1}%
{2}m\omega_{z}^{2}z^{2}$ \cite{Huang1}, is given by%
\begin{equation}
\frac{\mu_{\bot}(z)}{\hbar\omega_{\bot}}=(|q|+1)+\frac{1}{2}(\sqrt{\lambda
}\,\overline{Z})^{2}\left(  1-\frac{z^{2}}{Z^{2}}\right)  .\label{eq27}%
\end{equation}
The above analytical formulas generalize those obtained in Ref. \cite{Previo}
to the case of condensates containing an axisymmetric vortex. This is
particularly interesting because, in this case, the usual TF approximation
does not lead to explicit analytical formulas for the condensate properties.
In the absence of vortices $\beta_{q}\rightarrow1$ and one recovers the
results of Ref. \cite{Previo}.

Substituting Eq. (\ref{eq27}) into Eq. (\ref{eq26}) and solving for
$\overline{\mu}_{\bot}(z)\equiv\mu_{\bot}(z)/\hbar\omega_{\bot}$ one obtains
the local chemical potential as a function of the condensate density per unit
length%
\begin{equation}
\overline{\mu}_{\bot}(z)=(|q|+1)+\sqrt{\beta_{q}^{2}+4an_{1}(z)}-\beta_{q}.
\label{eq28}%
\end{equation}
This equation in the absence of vortices takes the simple form%
\begin{equation}
\overline{\mu}_{\bot}(z)=\sqrt{1+4an_{1}(z)}. \label{eq29}%
\end{equation}
In the TF regime ($\chi_{1}\gg1\rightarrow an_{1}\gg1$) the above expression
reduces to%
\begin{equation}
\overline{\mu}_{\bot}(z)=2\sqrt{an_{1}(z)}, \label{eq30}%
\end{equation}
which is the well-known result that can obtained directly from the GPE in this
regime \cite{Strin1}. In the mean-field quasi-1D limit ($\chi_{1}%
\ll1\rightarrow an_{1}\ll1$) Eq. (\ref{eq29}) reads%
\begin{equation}
\overline{\mu}_{\bot}(z)=1+2an_{1}(z). \label{eq31}%
\end{equation}
This is again the correct result that follows from the perturbative solution
of the GPE in this limit \cite{Strin1}. In fact, it can be easily verify that
all the analytical formulas derived in this section have the correct limits in
both the TF and the perturbative regimes \cite{Previo}.

From the above results one can derive a general formula for the (local) axial
(first) sound velocity $c_{\mathrm{1D}}$ of a cigar-shaped condensate, defined
by%
\begin{equation}
c_{\mathrm{1D}}^{2}=\frac{n_{1}}{m}\frac{\partial\mu_{\bot}}{\partial n_{1}%
}.\label{eq32}%
\end{equation}
Substitution of Eq. (\ref{eq28}) in Eq. (\ref{eq32}) leads to%
\begin{equation}
\frac{mc_{\mathrm{1D}}^{2}}{\hbar\omega_{\bot}}=\sqrt{\frac{4a^{2}n_{1}%
^{2}(z)}{\beta_{q}^{2}+4an_{1}(z)}}.\label{eq33}%
\end{equation}

In the absence of vortices and in the TF regime ($\beta_{q}=1$, $\chi_{1}\gg
1$) the expression above reduces to%
\begin{equation}
\frac{mc_{\mathrm{1D}}^{2}}{\hbar\omega_{\bot}}=\sqrt{an_{1}(z)}=\frac{1}%
{2}\overline{\mu}_{\bot}(z). \label{eq34}%
\end{equation}
This result is in agreement with the formula obtained by Zaremba for the sound
velocity of a homogeneous cigar-shaped condensate in the TF regime
\cite{Zaremba1}.

In the quasi-1D mean field regime with no vortices ($\chi_{1}\ll1$, $\beta
_{q}=1$) Eq. (\ref{eq33}) reduces to%
\begin{equation}
\frac{mc_{\mathrm{1D}}^{2}}{\hbar\omega_{\bot}}=2an_{1}(z)=\overline{\mu
}_{\bot}(z)-1, \label{eq35}%
\end{equation}
which is also the correct result as follows from the substitution of Eq.
(\ref{eq31}) into Eq. (\ref{eq32}).

For the particular case of a homogeneous cigar-shaped condensate (no axial
confinement and $n_{1}$ constant) the authors of Ref. \cite{Reatto2} have
obtained an analytical expression for the axial sound velocity that reproduces
correctly the perturbative result and, to a good approximation (with a
relative error less than 3\%), also the TF result, interpolating between these
two regimes. This expression has been obtained using an approach different
from that used here and, in fact, it exhibits a functional dependence on
$an_{1}$ that is very different from that in Eq. (\ref{eq33}). However,
particularizing Eq. (\ref{eq33}) for an axially homogeneous condensate one
finds that, in this case, both expressions are in quantitative agreement
within 3.75\%.%

%TCIMACRO{\FRAME{ftbpFU}{8.4076cm}{9.7614cm}{0pt}{\Qcb{(Color online)
%Theoretical prediction for the ground-state properties (in units of
%$\hbar\omega_{z}$) of arbitrary condensates with $q=0$ in different trap
%geometries $\lambda$ (solid lines). The open circles are the exact numerical
%results.}}{\Qlb{Fig1}}{Fig1.eps}{\special{ language "Scientific Word";
%type "GRAPHIC";  maintain-aspect-ratio TRUE;  display "USEDEF";
%valid_file "F";  width 8.4076cm;  height 9.7614cm;  depth 0pt;
%original-width 7.3408in;  original-height 8.5338in;  cropleft "0";
%filename 'Fig1.eps';file-properties "XNPEU";}}}%
%BeginExpansion
\begin{figure}
[ptb]
\begin{center}
\includegraphics[
height=9.7614cm,
width=8.4076cm
]%
{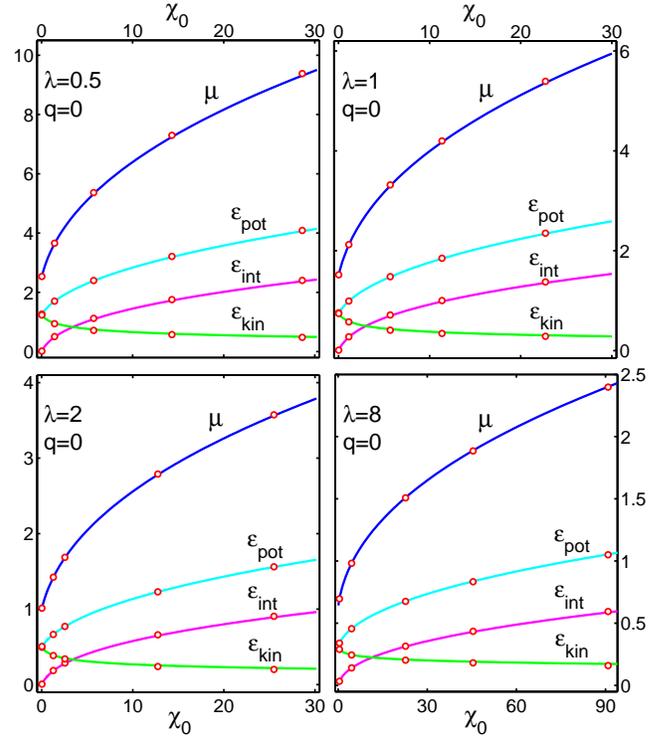}%
\caption{(Color online) Theoretical prediction for the ground-state properties
(in units of $\hbar\omega_{z}$) of arbitrary condensates with $q=0$ in
different trap geometries $\lambda$ (solid lines). The open circles are the
exact numerical results.}%
\label{Fig1}%
\end{center}
\end{figure}
%EndExpansion

\section{IV. NUMERICAL RESULTS}

To verify the predictions of our model we have numerically solved the
stationary Gross-Pitaevskii equation (\ref{eqGP}) by using a pseudospectral
method evolving in imaginary time. Figure \ref{Fig1} shows the ground-state
properties of condensates with $q=0$ and an arbitrary number of particles
($\chi_{0}$) in different trap geometries ($\lambda$). The open circles are
exact numerical results obtained from the Gross-Pitaevskii equation. The solid
lines are the theoretical predictions (in units of $\hbar\omega_{z}$) obtained
from Eqs. (\ref{eq15})--(\ref{eq20}). We have solved the polynomial equation
(\ref{eq16})\ by using a symbolic software package. As already said, this is a
trivial computational task that only requires one to type in a single instruction.

As is evident from Fig. \ref{Fig1}, the agreement is very good in all trap
geometries (typically better than $1\%$). In particular, although our
formalism is cylindrically symmetric it can describe accurately the properties
of spherical condensates ($\lambda=1$). For trap anisotropies higher than
those considered in Fig. \ref{Fig1} one can make use of the analytical
solutions (\ref{eq22}) and (\ref{eq24b}) found above. These cases will be
examined below (Fig. \ref{Fig5}).%
%TCIMACRO{\FRAME{ftbpFU}{8.4076cm}{9.7909cm}{0pt}{\Qcb{(Color online)
%Theoretical prediction for the ground-state properties (in units of
%$\hbar\omega_{z}$) of arbitrary condensates with a $q=1$ vortex in different
%trap geometries $\lambda$ (solid lines). The open circles are the exact
%numerical results.}}{\Qlb{Fig2}}{Fig2.eps}%
%width 8.4076cm;  height 9.7909cm;  depth 0pt;  original-width 7.3408in;
%original-height 8.5612in;  cropleft "0";  croptop "1";  cropright "1";
%cropbottom "0";  filename 'Fig2.eps';file-properties "XNPEU";}}}%
%BeginExpansion
\begin{figure}
[ptb]
\begin{center}
\includegraphics[
height=9.7909cm,
width=8.4076cm
]%
{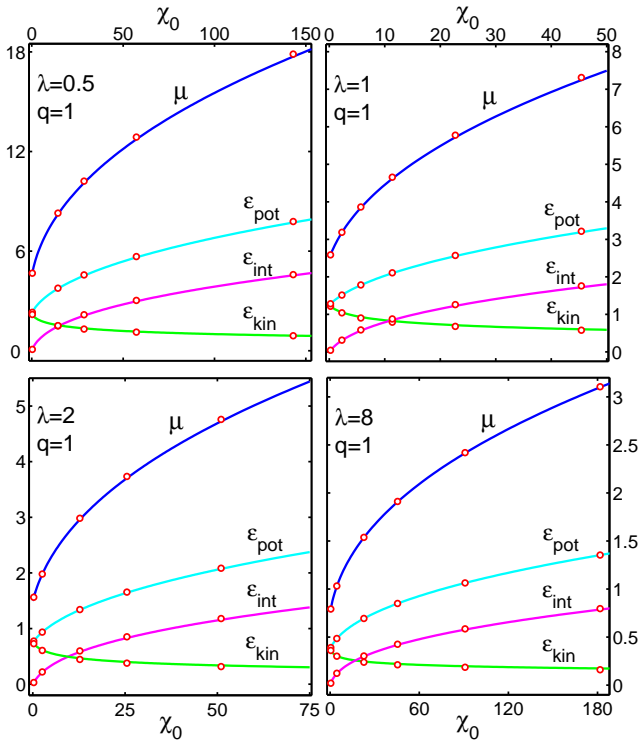}%
\caption{(Color online) Theoretical prediction for the ground-state properties
(in units of $\hbar\omega_{z}$) of arbitrary condensates with a $q=1$ vortex
in different trap geometries $\lambda$ (solid lines). The open circles are the
exact numerical results.}%
\label{Fig2}%
\end{center}
\end{figure}
%EndExpansion%
%TCIMACRO{\FRAME{ftbpFU}{8.4076cm}{9.8479cm}{0pt}{\Qcb{(Color online) Same as
%Fig. \ref{Fig2} for arbitrary condensates with a $q=2$ vortex.}}{\Qlb{Fig3}%
%}{Fig3.eps}{\special{ language "Scientific Word";  type "GRAPHIC";
%width 8.4076cm;  height 9.8479cm;  depth 0pt;  original-width 7.2876in;
%original-height 8.5454in;  cropleft "0";  croptop "1";  cropright "1";
%cropbottom "0";  filename 'Fig3.eps';file-properties "XNPEU";}}}%
%BeginExpansion
\begin{figure}
[ptb]
\begin{center}
\includegraphics[
height=9.8479cm,
width=8.4076cm
]%
{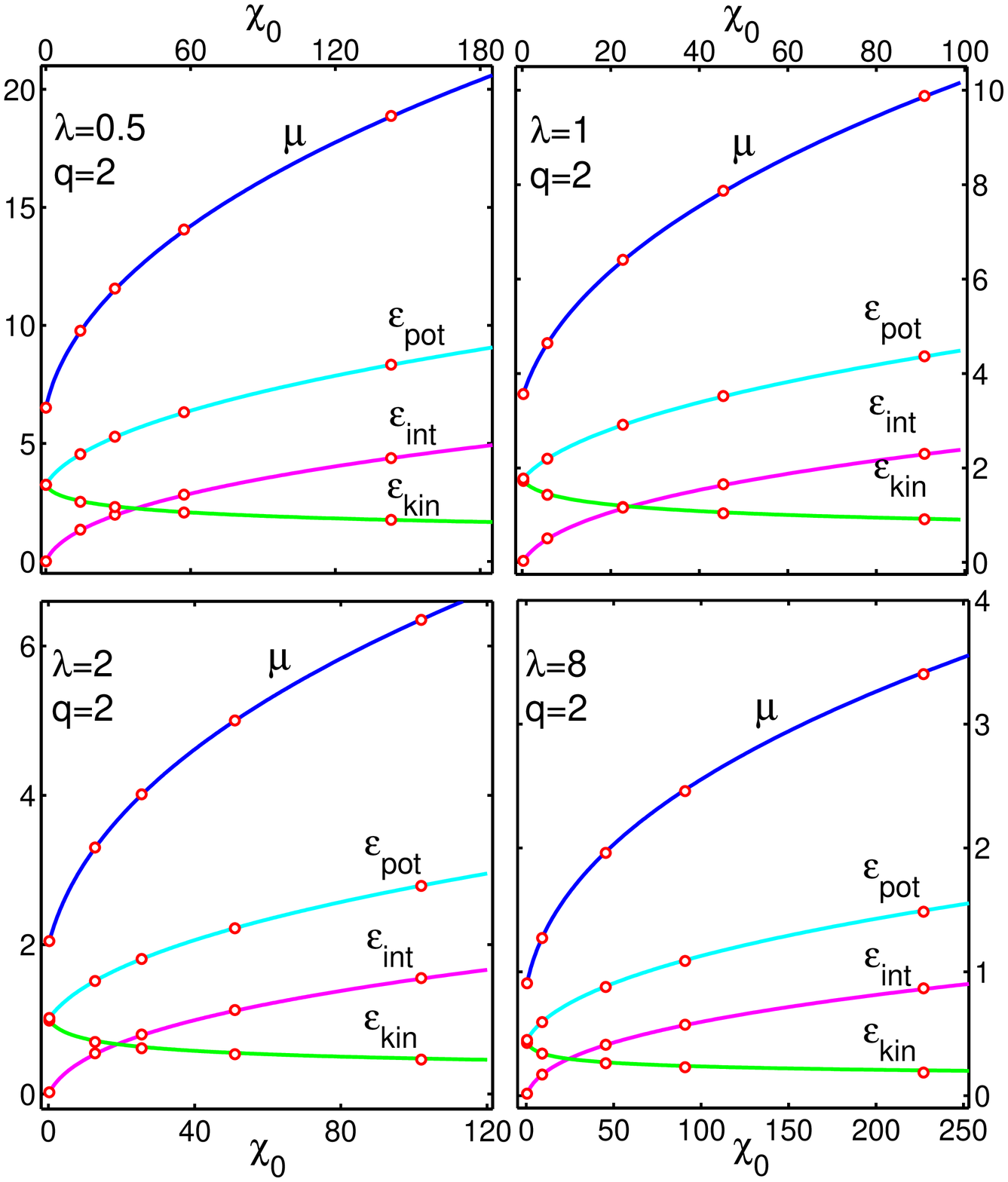}%
\caption{(Color online) Same as Fig. \ref{Fig2} for arbitrary condensates with
a $q=2$ vortex.}%
\label{Fig3}%
\end{center}
\end{figure}
%EndExpansion%
%TCIMACRO{\FRAME{ftbpFU}{8.4076cm}{9.8015cm}{0pt}{\Qcb{(Color online) Same as
%Fig. \ref{Fig2} for arbitrary condensates with a $q=4$ vortex.}}{\Qlb{Fig4}%
%}{Fig4.eps}{\special{ language "Scientific Word";  type "GRAPHIC";
%width 8.4076cm;  height 9.8015cm;  depth 0pt;  original-width 7.2876in;
%original-height 8.5072in;  cropleft "0";  croptop "1";  cropright "1";
%cropbottom "0";  filename 'Fig4.eps';file-properties "XNPEU";}}}%
%BeginExpansion
\begin{figure}
[ptb]
\begin{center}
\includegraphics[
height=9.8015cm,
width=8.4076cm
]%
{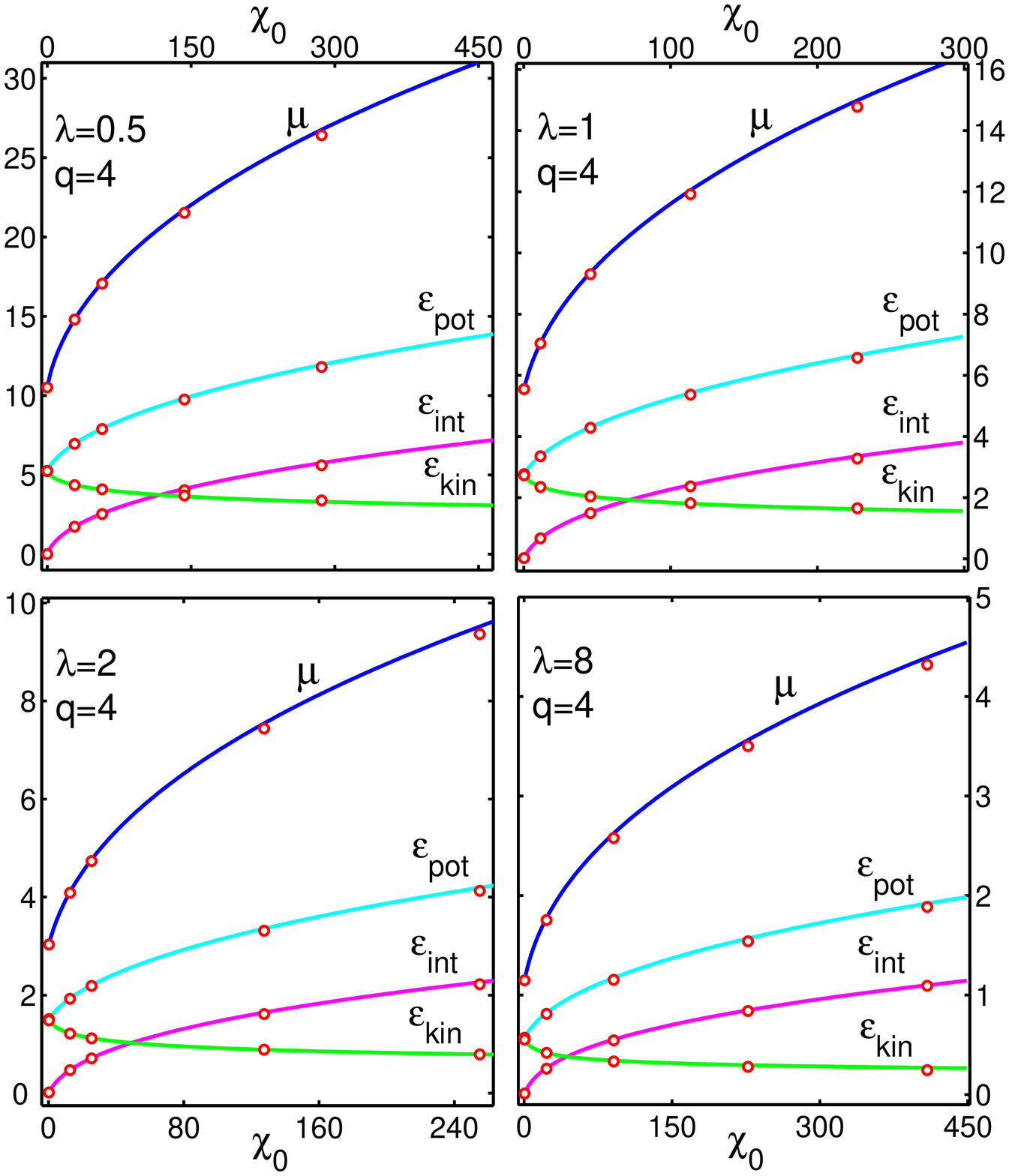}%
\caption{(Color online) Same as Fig. \ref{Fig2} for arbitrary condensates with
a $q=4$ vortex.}%
\label{Fig4}%
\end{center}
\end{figure}
%EndExpansion

In Figs. \ref{Fig2}--\ref{Fig4} we show the ground-state properties of
condensates containing an axisymmetric vortex of charge $q=1$, $2$, and $4$,
respectively. As before, the solid lines are the theoretical results obtained
from Eqs. (\ref{eq15})--(\ref{eq20}). These figures show that, regardless of
the number of particles ($\chi_{0}$) and trap geometry ($\lambda$), our model,
despite its simplicity, can also reproduce very accurately the physical
properties of condensates containing an axisymmetric vortex. This is
remarkable because, rather crudely, we have incorporated the effect of the
vortex into a uniform mean density $\bar{n}_{2}$ (with a renormalized
interaction strength $g\kappa_{1}^{-1}$) localized in the inner cylinder
$\,r_{\bot}\leq r_{\bot}^{0}$. In turn, the radius $r_{\bot}^{0}$ follows from
Eq. (\ref{eq6}), which cannot account for the effect of the mean-field
interaction energy. However, for condensates of intermediate size, the
interaction energy is no longer negligible in comparison with the kinetic
energy. In fact, it plays an important role in determining the size of the
vortex core, which decreases as $N$ increases. Equation (\ref{eq6}) cannot
incorporate this correction, which is proportionally more important for a high
$q$ and an intermediate number of particles (for a sufficiently large $N$ the
overall contribution of the vortex becomes negligible). Thus, one expects the
formulas above to be less accurate\ as $q$ increases. This can be appreciated
from Fig. \ref{Fig4}, which shows that for condensates with a $q=4$ vortex and
a $\chi_{0}$ sufficiently large that $\epsilon_{\mathrm{int}}>\epsilon
_{\mathrm{kin}}$ the theoretical predictions are slightly less accurate than
those corresponding to a $q=1$ or a $q=2$ vortex.%
%TCIMACRO{\FRAME{ftbpFU}{8.3591cm}{9.987cm}{0pt}{\Qcb{(Color online) (a)-(b):
%Theoretical prediction for the ground-state properties (in units of
%$\hbar\omega_{\bot}$) of arbitrary cigar-shaped condensates with $\lambda\ll1$
%and different vortex charges $q$ (solid lines). The open circles are exact
%numerical results obtained with $\lambda=0.1$. (c)-(d): Theoretical prediction
%for the ground-state properties (in units of $\hbar\omega_{z}$) of arbitrary
%disk-shaped condensates with $\lambda\gg2(|q|+1)$ and different vortex charges
%$q$ (solid lines). The open symbols are exact numerical results obtained with
%$\lambda=100$.}}{\Qlb{Fig5}}{Fig5.eps}{\special{ language "Scientific Word";
%type "GRAPHIC";  maintain-aspect-ratio TRUE;  display "USEDEF";
%valid_file "F";  width 8.3591cm;  height 9.987cm;  depth 0pt;
%original-width 7.1789in;  original-height 8.5869in;  cropleft "0";
%filename 'Fig5.eps';file-properties "XNPEU";}}}%
%BeginExpansion
\begin{figure}
[ptb]
\begin{center}
\includegraphics[
height=9.987cm,
width=8.3591cm
]%
{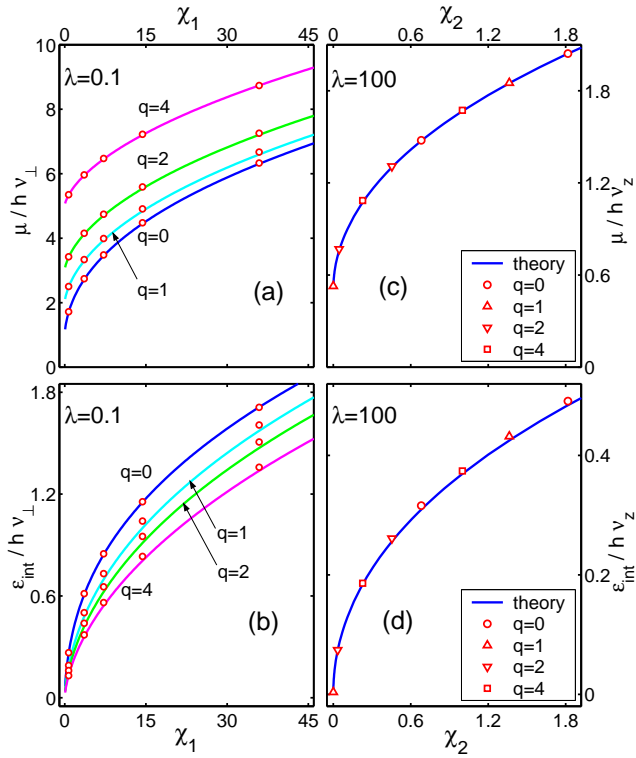}%
\caption{(Color online) (a)-(b): Theoretical prediction for the ground-state
properties (in units of $\hbar\omega_{\bot}$) of arbitrary cigar-shaped
condensates with $\lambda\ll1$ and different vortex charges $q$ (solid lines).
The open circles are exact numerical results obtained with $\lambda=0.1$.
(c)-(d): Theoretical prediction for the ground-state properties (in units of
$\hbar\omega_{z}$) of arbitrary disk-shaped condensates with $\lambda
\gg2(|q|+1)$ and different vortex charges $q$ (solid lines). The open symbols
are exact numerical results obtained with $\lambda=100$.}%
\label{Fig5}%
\end{center}
\end{figure}
%EndExpansion%
%TCIMACRO{\FRAME{ftbpFU}{8.2621cm}{6.5751cm}{0pt}{\Qcb{(Color online) (a):
%Theoretical prediction for the condensate density per unit length $n_{1}(z)$
%of arbitrary $\chi_{1}=1$ cigar-shaped condensates with $\lambda\ll1$ and
%different vortex charges $q$ (solid lines). The open symbols are exact
%numerical results obtained with $\lambda=0.1$. (b): Theoretical prediction for
%the condensate density per unit area $n_{2}(r_{\bot})$ of arbitrary $\chi
%_{2}=1$ disk-shaped condensates with $\lambda\gg2(|q|+1)$ and different vortex
%charges $q$ (solid line). The open symbols are exact numerical results
%obtained with $\lambda=100$.}}{\Qlb{Fig6}}{Fig6.eps}%
%width 8.2621cm;  height 6.5751cm;  depth 0pt;  original-width 7.1482in;
%original-height 5.6762in;  cropleft "0";  croptop "1";  cropright "1";
%cropbottom "0";  filename 'Fig6.eps';file-properties "XNPEU";}}}%
%BeginExpansion
\begin{figure}
[ptb]
\begin{center}
\includegraphics[
height=6.5751cm,
width=8.2621cm
]%
{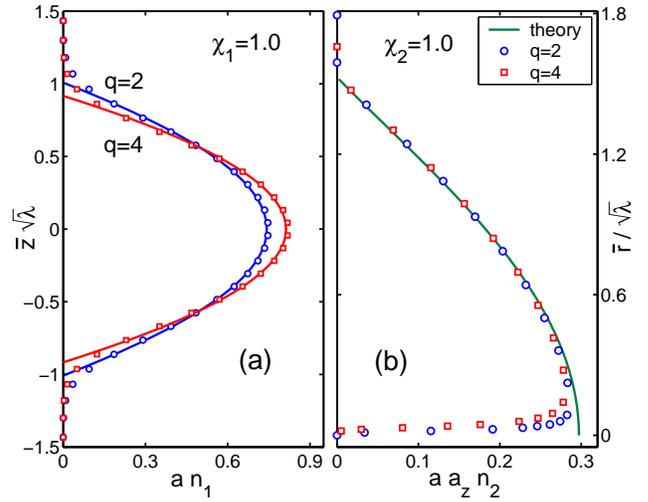}%
\caption{(Color online) (a): Theoretical prediction for the condensate density
per unit length $n_{1}(z)$ of arbitrary $\chi_{1}=1$ cigar-shaped condensates
with $\lambda\ll1$ and different vortex charges $q$ (solid lines). The open
symbols are exact numerical results obtained with $\lambda=0.1$. (b):
Theoretical prediction for the condensate density per unit area $n_{2}%
(r_{\bot})$ of arbitrary $\chi_{2}=1$ disk-shaped condensates with $\lambda
\gg2(|q|+1)$ and different vortex charges $q$ (solid line). The open symbols
are exact numerical results obtained with $\lambda=100$.}%
\label{Fig6}%
\end{center}
\end{figure}
%EndExpansion

Next, we consider the ground-state properties of condensates confined in
disk-shaped traps satisfying $\lambda\gg2(|q|+1)$ and cigar-shaped traps
satisfying $\lambda\ll1$. As already seen, in these limiting cases the
theoretical results become independent of $\lambda$ and can be directly
obtained from explicit analytical formulas. In Fig. \ref{Fig5} we show the
chemical potentials and interaction energies obtained from Eqs. (\ref{eq22}%
)--(\ref{eq23}) and (\ref{eq24b})--(\ref{eq25}), along with exact numerical
results. The corresponding kinetic and potential energies follow immediately
from Eqs. (\ref{eq20}).

As Fig. \ref{Fig5}(a) reflects, the chemical potential increases with the
vortex charge, a consequence of the larger kinetic and potential energies
associated with multiply quantized vortex states. However, as shown in Fig.
\ref{Fig5}(b), the opposite occurs with the mean interaction energy, which
decreases as $q$ increases because of the dilution effect that the centrifugal
barrier produced by the vortex has on the mean condensate density. The small
errors that can be appreciated in Fig. \ref{Fig5}(b) are due, in part, to the
fact that, for $q\neq0$, the approximate solution (\ref{eq24b}) incorporates a
residual error of order $2$--$3$ $\%$. The exact solution of the polynomial
equation (\ref{eq24}) leads to somewhat better results.

On the other hand, Figs. \ref{Fig5}(c) and \ref{Fig5}(d) reflect the fact that
for highly asymmetric trap geometries satisfying $\lambda\gg2(|q|+1)$ the
vortex contribution to the condensate properties becomes negligible.

Figure \ref{Fig6}(a) shows the condensate density per unit length $n_{1}(z)$
of arbitrary cigar-shaped condensates with $\lambda\ll1$ [obtained from our
analytical formulas (\ref{eq24b}) and (\ref{eq26})], while Fig. \ref{Fig6}(b)
shows the condensate density per unit area $n_{2}(r_{\bot})$ of arbitrary
disk-shaped condensates with $\lambda\gg2(|q|+1)$ [obtained from Eqs.
(\ref{eq22}) and (\ref{eq23b})].

The good agreement between theoretical and exact results demonstrates that the
formulas derived above are applicable for any trap geometry.

\section{V. CONCLUSION}

In a previous work we derived very accurate approximate analytical expressions
for the ground-state properties of trapped spherical, cigar-shaped, and
disk-shaped condensates with an arbitrary number of atoms in the mean-field
regime \cite{Previo}. In this work we have extended our previous proposal and
have derived general approximate formulas that provide with remarkable
accuracy the ground-state properties of any mean-field scalar Bose-Einstein
condensate with short-range repulsive interatomic interactions, confined in
arbitrary cylindrically symmetric harmonic traps, and even containing a
multiply quantized axisymmetric vortex.

In the appropriate limits (corresponding to cigar-shaped and disk-shaped
condensates) the ground-state properties follow from explicit analytical
formulas that generalize those obtained in Ref. \cite{Previo}. In the general
case, however, one has to solve a quintic polynomial equation. In this regard,
it is important to note that while solving the GP equation can be a complex
computational problem (especially in highly asymmetric trap geometries),
solving a polynomial equation is a trivial computational task. Using\ the
built-in capabilities of symbolic software packages such as
\textsc{mathematica} or \textsc{matlab} one obtains an instantaneous result
after typing in a single instruction.

The model presented in this work is essentially a convenient approximation
method motivated by two simple ideas: (i) There exists a direct relation
between the number of particles and the size of a trapped BEC and (ii) the
contribution from the harmonic oscillator energy to the chemical potential
cannot be smaller than the zero-point energy. Applying these simple ideas one
finds useful formulas of great generality that provide the condensate
ground-state properties in terms of the correct physically relevant
magnitudes. Moreover, even though no freely adjustable parameters are
introduced, in all cases the formulas obtained reproduce simultaneously with a
remarkable accuracy the condensate chemical potential and the interaction
energy and, as a consequence, the kinetic and the potential energy as well.
And this is true for mean-field condensates with any number of particles,
confined in any axisymmetric harmonic trap, and even containing an
axisymmetric vortex.

Finally, note that the results of this work are applicable in general to any
nonlinear system characterized by the stationary nonlinear Schr\"{o}dinger
equation (\ref{eqGP}).

\begin{acknowledgments}
This work has been supported by MEC (Spain) and FEDER fund (EU) (Contract No. Fis2005-02886).
\end{acknowledgments}

\bigskip

\end{document}